\documentstyle[12pt]{article}
\newcommand{\newsection}[1]{
\addtocounter{section}{1}
\setcounter{equation}{0}
\setcounter{subsection}{0}
\addcontentsline{toc}{section}{\protect
\numberline{\arabic{section}}{{\rm #1}}}
\vglue .6cm
\pagebreak[3]
\noindent{\bf  \thesection. #1}\nopagebreak[4]\par\vskip .3cm}
\newcommand{\newsubsection}[1]{
\addtocounter{subsection}{1}
\addcontentsline{toc}{subsection}{\protect
\numberline{\arabic{section}.\arabic{subsection}}{#1}}
\vglue .4cm
\pagebreak[3]
\noindent{\it \thesubsection. #1}\nopagebreak[4]\par\vskip .3cm}

\renewcommand{\theequation}{\thesection.\arabic{equation}}

\newcommand{\ben}{\begin{enumerate}}
\newcommand{\een}{\end{enumerate}}
\newlength{\extraspace}
\newlength{\extraspaces}
\newcounter{dummy}
\newcommand{\bc}{\begin{center}}
\newcommand{\ec}{\end{center}}
\newcommand{\be}{\begin{equation}
\addtolength{\abovedisplayskip}{\extraspaces}
\addtolength{\belowdisplayskip}{\extraspaces}
\addtolength{\abovedisplayshortskip}{\extraspace}
\addtolength{\belowdisplayshortskip}{\extraspace}}
\newcommand{\ee}{\end{equation}}

\newcommand{\ba}{\begin{eqnarray}
\addtolength{\abovedisplayskip}{\extraspaces}
\addtolength{\belowdisplayskip}{\extraspaces}
\addtolength{\abovedisplayshortskip}{\extraspace}
\addtolength{\belowdisplayshortskip}{\extraspace}}
\newcommand{\ea}{\end{eqnarray}}

\newcommand{\ban}{\begin{eqnarray*}
\addtolength{\abovedisplayskip}{\extraspaces}
\addtolength{\belowdisplayskip}{\extraspaces}
\addtolength{\abovedisplayshortskip}{\extraspace}
\addtolength{\belowdisplayshortskip}{\extraspace}}
\newcommand{\ean}{\end{eqnarray*}}
\newcommand{\baa}{                         
\addtocounter{equation}{1}
\setcounter{dummy}{\value{equation}}
\setcounter{equation}{0}
\renewcommand{\theequation}{\thesection.\arabic{dummy}\alph{equation}}
\begin{eqnarray}
\addtolength{\abovedisplayskip}{\extraspaces}
\addtolength{\belowdisplayskip}{\extraspaces}
\addtolength{\abovedisplayshortskip}{\extraspace}
\addtolength{\belowdisplayshortskip}{\extraspace}}
\newcommand{\eaa}{                                       
\end{eqnarray}
\setcounter{equation}{\value{dummy}}
\renewcommand{\theequation}{\thesection.\arabic{equation}}}

\input epsf

\newcounter{fignum}
\newcounter{tabel}

\newcounter{tabnum}
\newcounter{xxx}
\newcommand{\bl}{\begin{list}{({\it\roman{xxx}})}{\usecounter{xxx}}}
\newcommand{\el}{\end{list}}
\newcommand{\ppt}[1]{{\partial \over \partial t}}            
\newcommand{\ppx}[1]{{\partial \over \partial x}}            
\newcommand{\pqt}[1]{{\partial^2 \over \partial t^2}}            
\newcommand{\pqx}[1]{{\partial^2  \over \partial x^2}}            
\def\a{\alpha}

\def\k{\kappa}

\def\m{\mu}

\def\<{\langle}
\def\>{\rangle}

\newfont{\gothic}{eufm10 scaled\magstep1}

\renewcommand{\hat}{\widehat}

\begin{document}
\begin{titlepage}
\begin{flushleft}
\today
\end{flushleft}
\begin{flushright}
%This is the preprint no
\vbox{UCU-SCI-02-02}
\end{flushright}
\vskip 5cm
\begin{center}
{\Large\bf On Pay-off induced Quantum Games }
\vskip 1cm
\mbox{F.M.C.\ Witte}\\
{
\it
Department of Sciences, 
University College Utrecht\\
P.O. Box 80145, 3508 TC, Utrecht\\
Netherlands}
\end{center}
\vskip 1.5cm
\begin{abstract}
In recent years methods have been proposed to extend classical game theory 
into the quantum domain. This paper explores further extensions of 
these ideas that may have a substantial potential for further research. 
Upon reformulating quantum game theory as a theory of classical 
games played by "quantum players" I take a constructive approach.
The roles of the players and the arbiter are investigated for 
clues on the nature of the quantum game space.  

Upon examination of the role of the arbiter, a possible non-commutative 
nature of pay-off operators can be deduced. I investigate a sub-class 
of games in which the pay-off operators satisfy non-trivial commutation 
relations. Non-abelian pay-off operators can be used to generate whole 
families of quantum games.
\end{abstract}
\end{titlepage}
\newpage

\newsection{Introduction}
Classical Game theory is a subject that has been under substantial study ever
since its origins in the book of von Neumann and Morgenstern \cite{Neumann}.
A crucial step forward in the understanding of the dynamics of games was made 
by John Nash \cite{Nash} upon the introduction of well-defined equilibrium states,
the so-called Nash equilibrium. It is however well known that not all games posses
a Nash equilibrium rendering them essentially unsolvable.

A number of papers have appeared in which it is attempted to quantize
classical games \cite{earlyQG}. Essentially a classical game is quantized by 
assuming that the possible strategies a player can choose are elements if a 
Hilbert space of strategies that also accomodates superpositions of strategies.
The players can apply unitary operations to the strategies, called "tactics",
yielding a "final state" that then forms the basis for the determination of 
the pay-off. A physical model that guides these considerations is that of 
"classical" players communicating their strategies to the referee by means 
of "quantum objects". One difficulty with this type of quantization of a classical 
game is the appearance of an "initial strategy" which has no classical equivalent
and no particular meaning. This is one of the problems the present paper attempts to 
resolve.

In this paper I will describe an extension of the recently proposed 
methods for quantizing classical games. First of all, I will reformulate quantum
game theory in terms of quantum players playing a classical game. This then 
not only incorporates, but also gives meaning to the notion of an "initial strategy".
It opens up he way to acknowledging that the origin and scale of quantum 
phenomena in games not neccesarilly in causal correspondence with the quantum 
mechanics that underlies physical nature.

The notion of a player is coupled to the existence of a corresponding pay-off 
operator, rather than a particular subspace of the quantum game Hilbert space. 
Conceivably one special class of games is that generated my non-commuting pay-off 
operators. This gives rise to an esentially new concept in quantum game theory. 
Non-commutation induces a intrinsic uncertainty in a player's pay-off. As examples
I will discuss a simple 2-player game and a set of $U(2)$ generated quantum games.

The material in this paper is ordered as follows. 
In the second section I will discus the quantization of games, contrasting the 
recipe of quantizing players against that of quantizing strategies. 
In a short subsection I will illustrate some of the concepts from section 
two for the special case of the quantum Battle of the Sexes.
In the third section I discuss non-commuting pay-off operators generating a 
quantum game and illustrate the theory by analysing the game content of the 
$U(2)$ pay-off algebras. In particular I will present arguments in favor of 
allowing pay-off operators to be generally non-commuting. Finally I will 
summarize the main conclusions.

\newsection{Classical and Quantum Games}
A classical game consists of a number of players $I_{j}$, a discrete set $\cal{S}$ of 
strategies $s_{i} \epsilon \cal{S}$ these players can adhere to, and a pay-off function 
$R(I_{j},s_{k})$ wich assigns a real number to each player depending on his own strategy 
and that of the others. Game theory is the study of how these ingredients give rise to a 
rational development of the game. 

A "solution" of the game would consist of a combination of strategies such that they 
will surely be played by rational players. A usefull tool in analyzing games is that 
of a Nash equilibrium. This essentially is a set of strategies which is such, that 
a change in strategy by a single player would result in a decrease of the players pay-off. 
Hence, for a player in a Nash equilibrium situation there is no rational incentive to 
alter his strategy. A problem with Nash equilibria is that they are not neccesarilly 
unique. Hence, if a given game allows for two distinct Nash equilibria the concept 
loses some of its utillity as a "solution" to the game. 

It is possible to extend the concept of strategies to include games in which a probabillity 
is asigned to each strategy. An expected pay-off can be computed and analyzed for the 
existence of apropriate Nash equilibria. Such a procedure can give rise to new equilibria 
in a game, though not neccesarilly with a higher pay-off for the players. The pay-off
determined in such cases is a strict statistical average of the basic pay-offs defining the 
game. The drawback of this is that the notion of players playing a game by randomly, but 
coherently, changing their strategies according to some probabillity distribution seems 
a situation that can hardly survive in complicated games. The succes of statistical mechanics 
basically relies on almost unreasonable effectiveness of the Bolzmannweight, which can be
traced to energy-conservation. No such unique and unreasonably effective probabillity 
distribution seems likely to exist for human players of a classical game.

In current formulations of Quantum Game Theory, the discrete set of strategies is replaced by 
a vectorspace with a norm related to the probabillity of a certain strategy being played. Pay-off
is represent through a set of hermitian operators. In a sense, the definition of the game requires
the specification of the eigenvalues of these operators. The origin of the quantization is often 
sought in the neccesity of the players to store the choice for a particular strategy in terms of 
observables of some quantum system. For example, a player having the choice between two possible 
strategies could choose express to his choice by manipulating an electron-spin. The referee 
measuring the spin-state of the electron thus is confronted with all aspects of the quantum 
measurement problems, such as the possibillity of superposed spin-states. However, it is 
rarely noted that the player would suffer from the same problem, that is an inabillity to 
actually produce a spin-state which properly reflects his choice of strategy. As a result the 
game becomes a game of chance on all hands. Nonetheless there is some virtue in quantizing games. 
It has been shown by Marinatto and Weber, for a particular set of two-player games, that for 
"entangled strategies" a unique Nash equilibrium arises that offers both players an optimal 
pay-off when contrasted to other equilibria. If we adhere to the belief that this is no
coincidence but rather a generic feature of quantum games, they seem to promise unique solutions. 
A conceptual problem that remains is that the concept of an entangled strategy 
cannot really be expressed into macroscopic terms. Some of the issues brought up 
in this section can be resolved by a mere shift in the interpretation of the origin 
of the quantization.

\newsubsection{Quantum Players}
The first new element that I seek to introduce into this discussion is that of a quantum player. 
The central idea is to accept the classical nature of the strategies to be played, but to adhere
the quantum nature of the quantum game to the decision-making process in each individual player.
To provide some motivation for this let me give two lines of reasoning now.

Any player in a game will play the same game "in his imagination" several times before coming up 
with a choice of strategy. This "virtual" part of the classical game is never taken into 
consideration, apart from discussing the extent to which the players are considered 
rational or not. 
In the view of this paper, the virtual game runs until the moment in which the arbiter 
requires the players to make their moves. The arbiter can then apply a measurement to 
observe the "strategic state" of each individual player. The moment at which he does 
so the sequence of virtual games is essentially interrupted. This view, reminiscent 
of the "sum over all histories", or path-integral quantization, argues for a {it 
diffusion-like} process, if the weights of the different virtual games are real 
and a {\it quantum-like} process if the weights are taken to be complex. It seems
natural to assume that the key notion leading to a complex weight of some kind, 
is the fact that the virtual games are never really played. If they were, they 
would simply be part of the game and as such could only account for the statistical 
fluctuations in an actually played game. 

The notion of not playing the virtual game, is similar to the notion of not "measuring" the
paths of the electrons in two-slit experiments. If we measure the electron paths the quantum 
interference effects dissappear and a statistical noise remains. Further support for
such a view should come from an actualy construction of such a sum over all histories in
a classical game. Infact, this point of view possibly offers a resolution of a debate
concerning the so-called theory of moves \cite{Moves} and the objections that have been
brought forward against it, for example in \cite{Nomoves}. It would however entail that
quantizing a classical game in some extent incorporates dynamical information into the 
states. We will see that this indeed appears to be the case for the non-commutative 
games.

A second way of regarding these matters is by considering the origin of the state-space
in which quantum games live. Strategies are rather unnnatural things to superimpose on 
one another. However, it seems conceivable that the players {\it preference} for a 
particular strategy may not be a discrete variable. Players could be assumed to be able
to experience a "superposition" of preferences. This should be distinguished from something
we might identify with "doubt"; a situation in which a player will tend to fluctuate
between two options being in favor of one a certain amount of the time and in favor of the
other at other times. The superposition of preferences conceivably reflects the process
of a player "not having made up his mind". Again, as the player is not choosing, there
is no actual measurement of his choice of strategy and concequently these fluctuations
cannot be taken to weighed by a real-valued weight.

Let me summarise the ingredients for the description of an N-player quantum game. 
I identify the following parts;
\bl
\item {\bf N-Player Game space:}  The state of the N-player game can be represented by
a vector in a suitable Hilbert space.
\item {\bf Pay-off:} There exists a set of N, linear, self-adjoint pay-off operators
on the N-player Hilbert space, one for every player present.
\item {\bf Arbiter:} There exists an arbiter who determines when the state of the game 
is measured, and determines pay-off.
\el

Some comments are in order here.

Essentially the set op pay-off operators defines the game both in terms of pay-off for
strategies, as well as in terms of the number of players. It is important to distinguish
between commutative and {\it non-commutative} games. In general classical games deal with 
commutative games. In a commutative game the order in which pay-off is determined is 
irrelevant. In a non-commutative game changing the order would give rise to changing values 
for the pay-off, and to uncertainties in the amount of pay-off a certain state will gain
a given player. This is an argument that may lead one to contemplate the conjecture that
quantum games intrinsically are games with strategic uncertainty.

In contrast to earlier works the notion of players manipulating their quantum
strategies from an initial state into a final state has not been included. As
the players are not expected to measure the state of the game, it seems rather
unnatural for them to know the exact outcome of the manipulations they are
supposed to make. Non-commutative games appear to include dynamics from the
start, the reason for that will become clear below. It seems reasonable however
to suggest that generically the generator of time-evolution will, in some form, 
be constructable from the player's pay-off operators. In particular, I choose
to view the arbiter as an intrinsic object whose operators are related to the
pay-off operators of the players, suggesting a reformulation of the extensive
form of the games as endogenous, advocated in \cite{Moves}.

\newsection{Non-commutative operators in Quantum games}

The entangled strategies, that play a central role in the issue of solvabillity of a game 
according to some authors, actually never generate the predicted pay-off, as every measurement 
of the referee produces the pay-off eigenvalues that initially defined the game. Thus it 
remains questionable to me whether a rational player could actually play such a strategy!
Again, here we may ponder whether the referee or arbiter may use additional operators to
establish the state of the game, such as products of the pay-off operators.
For an entangled strategy to be measurable at all, one would require a hermitian operator 
that has these strategies among its eigen states. 
Such operators naturally would not commute with the two pay-off operators defining the 
game. There seems no real reason to introduce such non-commuting operators into the game 
unless they could be associated with either the pay-off of another player in the game or
the arbiter.

So the role of the arbiter deserves further attention. In standard formulation the 
arbiter is confronted with the quantum measurement-problem and is a hybrid mix of 
an entity operating inside and outside the game. This is the quantum analogy of the
edogenous versus exogenous interpretation of the extensive form of games reffered to
earlier \cite{Moves,Nomoves}. I want to do away with that by implementing two rules;
\bl
\item the operators representing the players can be constructed from the arbiter's 
operators alone.
\item a "minimal arbiter" should be able to do two things: {\bf a:} 
"measure" the sequential number of moves the game has gone through, and 
{\bf b:} map the $(n-1)^{th}$ move states onto the $n^{th}$ move states.
\el
What the second item entails is the following. The quantum game space of a game is 
supposed to contain far more states than proposed in earlier works. Not only are 
all strategies to be included, but in principle every move or round of the game 
could come with a new set of strategies. There ought to be an operator that 
counts the number of moves, as well as an operator that "lifts" the game by 
one move. These two operators we will identify with a {\it minimal arbiter}.

Although allowing for non-commutative games seems a complication, the algebra of 
non-commutative pay-off is powerfull and can give direct clues on the 
structure of the N-player Game space. We will analyze this for a number of 
examples in the following section.

Let us summarize the main points. Quantized games hold the promise of providing 
solvabillity while at the same time expanding our pool of playable games. There
are indications that entangled states play an important role in providing these
solutions. However, measuring the system to be in an entangled state, would
require the natural presence of non-commuting operators in the game. But natural 
in the definition of the game are the pay-off operators.

If pay-off operators were to be non-commuting the correspondence between classical 
games and the direct-product states of the quantum games would breakdown at the level 
of simultaneous diagonalisation. Hence non-commutation among the pay-off operators 
could be interpreted as the source of "quantumness" of the game. 
An operator representing the arbiter can either be constructed from the pay-off 
operators of the players, or by assigning a pay-off-like operator to the arbiter 
which he uses to decide upon the state of the system. In that case the 
arbiter enters, effectively, as another player with a slightly different role.

\newsection{Non-Commutative Quantum Games}
The above elements in a formulation of quantum game theory focus attention on the 
game space, and interprete the vectors therein as representing the state of the players.
The possible importance of entangled states in quantum games makes it desirable 
to interprete the appearance of non-commuting operators. But in the next subsection
we will show non-commutativity of pay-off operators can be supported with far less 
esoteric argumentation.

Players are recognisable elements of a game because, and only they receive 
pay-off from their strategies. Hence, it should be sufficient to specify the pay-off
to fix the game. Classically this indeed is the proper situation. The classical arbiter 
is very much like the classical observer in physics; hardly part of the system. A proper
treatment of measurement in quantum mechanics suggests the incorporation of the observers 
measuring aparatus into the quantum descriptionof the system. As macroscopic measuring 
devices behave notoriously classically this is computationally hardly a feasably approach.
However, for quantum games we can do that for and it provides us with a powerfull extension
of quantum game theory.

\newsubsection{A simple non-commutative 2-Player Game}
Consider a single minimal arbiter, as defined earlier in this text. This arbiter is 
represented by two operators. First of all there is the operator $\hat{N}$ that 
counts the sequential number of the moves played by the players. Obviously it must be
a hermitian operator as it corresponds to an observable.
Secondly there is the raising operator $\hat{\a}_{+}$ that raises this sequential 
move-number by one unit. As the operator itself does not correspond to an observable
a priory, there is no need for it to be hermitian. In order for this to work, the 
two operators must satisfy the following commutation relation,
\be
[\hat{N},\hat{\a}_{+}] = \hat{\a}_{+} \ .
\ee
If we define  $\hat{\a}_{-}$ to be the hermitian conjugate of $\hat{\a}_{+}$,
\be
\hat{\a}_{-} = \hat{\a}_{+}^{\dagger}
\ee
we can write $\hat{N}$ in terms of the two as
\be
\hat{N} = \hat{\a}_{+} \hat{\a}_{-} \ ,
\ee
where the operators $\hat{\a}_{+}$ and $\hat{\a}_{-}$ satisfy
\be
[ \hat{\a}_{-},\hat{\a}_{+}] = 1 \ . 
\ee
From this last result it is obvious that these operators are {\it not} hermitian.
They are well known as the ladder-operators in the quantum mechanics of the
harmonic oscillator. It follows from our premisses that the following elements 
have been given;
a set of two pay-off operators $\pi_{i}$. The two ladder-operators are
supposed to be related to the pay-off operators of the players. For simplicity 
I assume that they are {\it linearly} related. Given the fact that the pay-off operators 
are hermitian, the two sole possible combinations are straightforward,
\be
\hat{\a}_{\pm} = \frac{1}{\sqrt{2}} [ \frac{1}{\k_{1}} \hat{\pi}_{1} 
\pm \frac{\imath }{\k_{2}} \hat{\pi}_{2} ] \ ,
\ee
where $\k_{j}$ are suitable units of pay-off for the corresponding player. 
Now with the ladder operators expressed in terms of the pay-off operators, 
it is straightforward to show they satisfy
\be
[\hat{\pi}_{1},\hat{\pi}_{2}] = \imath \k_{1} \k_{2} \ .
\ee
Let me remark here that due to the commutation relation above, the two pay-offs 
will satisfy an uncertainty relation of the Heisenberg type.
A full analysis of this particular game has been given elsewhere
\cite{AQB}.

Basically such a game could occur when the arbiter has posession
of a collection of instable nculei of species $A$ that can decay with equal 
probabillity into species $B$ {heads} and $C$ \{tales\}. Two players place a bet on
the occurrence of either heads or tales every time a decay occurs. The arbiter
does not know the number of rounds played, i.e. the number of decays that have
occured, untill the moment he measures it and counts the number of heads and tails
and determines the pay-off for each player. In a Schr\"odinger Cat like case,
superpositions of states representing a different number of played rounds can
occur. They turn this game, which classically is a probabillistic gambing game, 
into a game where correlations between the two players pay-off may occur. An
analysis of the correlations between the pay-off $\pi_{1}$ and $\pi_{2}$ reveals
that the quantum game can have winners and losers depending on the details of
the superposition of states with a different number of rounds played \cite{AQB}.

\newsubsection{Multi-game players}
The construction of the previous subsection can be easilly extended in the following fashion.
In the above example there was one unique way to go from one round of moves to the next. However
it is conceivable many realistic games are "multi-games" in the sense that between two rounds
of one game there is a round of another game. Infact, negotiations and bargaining \cite{Qbargain}
will often involve sequentially playing a move in one game whose resulting pay-off affects the
next move in another game. Such multi-game players would have pay-off operators for each corresponding
game sub-space of the total game space. We cannot be sure however that the total pay-off operator
for a given player is always reducible to a direct product of pay-off operators in sub-spaces of
the total game space. What we can safely assume though, is that the arbiter is able to count the
number of moves in every sub-game as well as is able to increase the move-count by one.

Therefor defining a game played by multi-game players would start out by collecting an appropriate
set of $K$ counting operators $\hat{N}_{i}$ and of step-up operators $\hat{\a_{+ i}}$, where $i = 1 
... K$. They should satisfy the usual commutation relations given above,
\be
[\hat{N}_{i},\hat{\a}_{+ i}] = \hat{\a}_{+ i} \ ,
\ee
and
\be
[ \hat{\a}_{- i},\hat{\a}_{+ i}] = 1 \ . 
\ee
The question is whether we can say anything about the commutation relations of the step-up 
operators among one another. The answer is affirmative for a particular class of games, namely
those in which the numbercount of the number of rounds played in the sub-games is unambiguous.
In that case we should have
\be
[\hat{N}_{i},\hat{N}_{j}] = 0 \ .
\ee
The remainder of this paper will be devoted making a start with analyzing the game content of
these relations.

If we require a linear relationship between the step operators and the corresponding 
pay-off operators of the players we will find the game already analyzed above, but now 
generalized to $K$ pairs of 2 players. Only pairwise do they have non-commuting pay-off. 
Although we really have $K$ identical copies of the 2 player game, correlations may exist 
between all four pay-off operators for special choices of superpositions. Such games 
allowing for a variable number of players can be constructed by using methods of second 
quantisation. This yields interesting results that will be published elsewhere \cite{2ndQGame}.

\newsection{$K=2$ is a non-commutative game}

If we allow for bilinear combinations in relating the step-operators with pay-off
operators we will get
\be
P^{A} = P_{a b}^{A}\a_{- a} \a_{+ b} \ ,
\ee
where a sum over lower-case $a,b$ is understood. The commutation relations between the
pay-off operators will then be detemined by those of the matrices $P^{A}_{a b}$.
Such a construction yields pay-off operators that, if we require that they 
leave the total number of rounds played invariant, i.e.
\be
[\sum_{j }\hat{N}_{i},\hat{P}_{A}] = 0  \ ,
\ee
will form a representation of a sub-algebras of some $su(n)$ Lie-algebra. However, it is 
similarly conceivable that the rules of the game require the that the number of rounds
of some subset of $m$ types of moves contribute negatively, and $k$ types contribute
positively. In this case we find a representation of the $su(k,m)$ algebra.

Because we have introduced $2 K$ operators to define the game, we would like to see 
{\it at least} $2 K$ players, or $2K -1$ players and the arbiter, so as to guarantee 
that the definition requires no additional operators to be introduced beyond those 
used for constructing pay-off. Thus the upper-case index $A$ should run atleast from 
$1$ to $2 K$. 
We know however from our previous construction that apart from a single groundstate, we 
will have a total number of $K$ states at the level of the first round. In this 
subspace, the group of unitary transformations is generated by $K^2 - 1$ 
traceless hermitian operators. We call into memory that linear combinations of 
hermitian operators with {\it real} coefficients are again hermitian and that commutators 
of hermitian operators are anti-hermitian. The set of pay-off operators has to be 
a subset of these generators, hence the $SU(K)$ commutation relations will in part
yield commutation relations among the pay-off operators. If the number of players
$2K$ exceeds or is equal to the number of available hermitian operators, $K^2 - 1$,
then the pay-off algebra will be determined by the $SU(K)$ lie algebra.

Consequently we would generally for
\be
K^2 - 1 < 2 K \ ,
\ee
we are definitly dealing with non-commutative games.The values of $K$ for which 
this occurs is $K=1$ and $K=2$. The $K=1$ game has been previously discussed. Our
argument however indicates that for $K=2$ we have another family of non-commutative
quantum games. 

\newsubsection{The $SU(2)$ and $SU(1,1)$ Quantum Games}
I will study two algebras for the pay-off operators in this subsection. I will start
out discussing the compact $su(2)$ algebra that will lead to finite dimensional game
spaces, followed by a brief overview of the $su(1,1)$ algebra that will generate infinite
dimensional game spaces in every round.

The relevant commutation relations for $su(2)$ are
\be
[\hat{\pi}_{1},\hat{\pi}_{2}] = \imath \hat{\pi}_{3} \ , [\hat{\pi}_{2},\hat{\pi}_{3}] = 
\imath \hat{\pi}_{1} \ ,
[\hat{\pi}_{3},\hat{\pi}_{1}] = \imath \hat{\pi}_{2} \ .
\ee
The game space generated by these commutation relations is a representation of the $SU(2)$ 
Lie algebra,familiar from the quantum mechanics of angular momentum. Note that no 
matter what the dimensionality of the actual representation is, the game by construction is 
played by four players, one of which is the arbiter represented by the only Casimir operator
\be
\hat{P} = \hat{\pi}_{1}^2 + \hat{\pi}_{2}^2 + \hat{\pi}_{3}^2 \ .
\ee 
This indeed follows from the relation of its eigenvalues to those of the counting operators.

We know from the $su(2)$ representation theory that the eigenvalues of $\hat{P}$, together
with those of one of the other pay-off operators will generate a complete set of states. 
We write
\be
\hat{P} | \k ,\m \rangle = \k (\k + 1)| \k ,\m \rangle  \ , 
\hat{\pi}_{3}| \k ,\m \rangle = \m | \k ,\m \rangle  \ .
\ee
The commutator algebra then prescribes for unitary representations that
the eigenvalues are restricted to
\be
\k = 0, \frac{1}{2}, 1, \frac{3}{2} ... \ , \  \m = -\k , -\k +1 , ... , \k -1, \k \ .
\ee
The remaining pay-off operators cannot be diagonalised. As in the $su(2)$ algebra no single
pay-off operator can be distinguished from any of the others our choice is arbitrary convention
and every other choice describes the same gamespace. 

The quantum number $\k$ can be related to the eigenvalues of the two counting operators
$\hat{N}_{i}$, one finds
\be
k = \frac{N_{1} + N_{2}}{2} \ .
\ee
The application of the step-up operators hence allows us to step through the subsequent 
irreducible representations of $su(2)$. The $\k=0$ singlet is the initial state of the
game. The $\k=\frac{1}{2}$ doublet is the two-dimensional spin-representation of $SU(2)$. 
The quantum numbers labelling this representation are
\be
\k = \frac{1}{2} \ , \ \m = -\frac{1}{2} , \frac{1}{2} \ .
\ee
As we see one of the players wins or loses, the other two players will have a vanishing
expectation value for their pay-off. One can view this sub-space of the total gamespace 
as allowing for the formation of a coalition of two players against the remaining player.
This is also suggested by the subspace having just two strategies and thus not supporting 
a free-player basis. 
For higher values of $\k$ the subspace of the game will have $2\k+1$ different states, i.e. 
strategies that can be played in that round. Assume for a moment that the game is in an
eigenstate of player 3 pay-off. 

If we consider the relative uncertainty of the pay-off of the players 1 and 2, by scaling 
these by $\frac{1}{\sqrt{\k(\k+1)}}$ we find that the role of non-commutation also scales 
by the same factor. Hence, for moderate values of $\m$, and large values of $\k$ the pay-off 
of the players will tend towards classical commutativity. The classical limit will appear as 
a game in which the sum of the squared pay-off of the players is fixed to a continuous sphere 
whose radius is set by the referee, i.e. increases with game-time. The analysis of such a 
game falls outside the scope of this paper. Note however that the amount of pay-off a player
has obtained after a given number of rounds is allways bounded and increasing one player's pay-off
will allways go at the expense of the pay-off of others. 

Now we will see a different set of commutation rules that lead to an unbounded game and hence will 
require a different intepretation. The $su(1,1)$ commutation relations are
\be
[\hat{\pi}_{1},\hat{\pi}_{2}] = - \imath \hat{\pi}_{3} \ , [\hat{\pi}_{2},\hat{\pi}_{3}] = 
\imath \hat{\pi}_{1} \ ,
[\hat{\pi}_{3},\hat{\pi}_{1}] = \imath \hat{\pi}_{2} \ .
\ee
They are almost identical to those of $su(2)$ apart from the different sign in the result
of the commutator $[\hat{\pi}_{1},\hat{\pi}_{2}]$. This has three significant consequences,
two of which I will address here. First of all, it singles out one particular player, here
player 3. Furthermore it leads to the existence of inequivalent representations of 
this algebra depending on which of the three pay-off operators is diagonalised. 
Finally, it leads to representations that are essentially unbounded. We will restrict our 
attention here to the representations that arise from diagonalizing $\hat{\pi}_{3}$ 
and the Casimir operator
\be
\hat{P_{1,1}} = - \hat{\pi}_{1}^2 - \hat{\pi}_{2}^2 + \hat{\pi}_{3}^2 \ .
\ee
As you see we are dealing with an indefinite metric. Infact this algebra is directly related to
the $2+1$-dimensional Lorentzgroup $SO(2,1)$. Lorentztransformations will preserve the metric,
so the sectors with positive and negative values for the Casimir-operator will be disjoint.
If we go through the standard procedure for finding unitary representations \cite{SimpleQ} we
find spaces in which the the eigenvalues of $\hat{\pi}_{3}$ differ by unit-steps and are either
bounded from above {\it or} bounded from below by a number $\k$ which determines the eigenvalue
of the Casimir operator through the relationship $\k (\k + 1)$. Hence the player 3 seems to act 
as a second arbiter that is counting a number of rounds played in a game played by the players 
1 and 2.

Finally let me remark here that even the $K=1$ game will support a $su(2)$ or $su(1,1)$ algebra
if we consider pay-off operators which are bi-linear in both the step-up as well as the step-down
operators.

\newsection{Discussion and Conclusion}
In this paper I have sought to make two distinct points. Firstly, a more philosophical point,
it to consider quantum games as classical games played by quantum players, rather than quantum 
games played by classical players. This limits the capacity of players to manipulate the quantum 
state of the game, just as a single electron cannot manipulate its own spin state. It gives
proper meaning to the initial state and requires the inclusion of dynamics as a natural ingredient
of the game space. The dynamics being generated by counting the number of rounds played results
in the introduction of the apropriate operators to do so.

Secondly I presented arguments to support the view that non-commuting pay-off is worth
investigating. I have shown how a minimalistic approach to constructing quantum games and their
dynamics in terms of rounds naturally leads to non-commutative pay-off operators. 
I have given a brief analysis of the simplest games that can be seen to arise in this way. 
Obviously, in view of the above it becomes interesting to study the game content of unitary 
representations of compact lie groups, which is a well known area of mathematics. As we have 
seen, a non-compact lie-group leads to an unbounded gamespace which is a generic feature. It 
remains to be investigated in which sense this introduces a certain "trivialisation" in the 
sense that unbounded directions would generally have to be identified with arbiters counting 
numbers of rounds.

\end{document}